\documentclass[alpha-refs]{wiley-article}

\usepackage{graphicx}
\usepackage[space]{grffile}
\usepackage{latexsym}
\usepackage{textcomp}
\usepackage{longtable}
\usepackage{tabulary}
\usepackage{booktabs,array,multirow}
\usepackage{amsfonts,amsmath,amssymb}
\usepackage{natbib}
\usepackage{url}
\usepackage{hyperref}
\hypersetup{colorlinks=false,pdfborder={0 0 0}}
\usepackage{etoolbox}
\newif\iflatexml\latexmlfalse

\AtBeginDocument{\DeclareGraphicsExtensions{.pdf,.PDF,.eps,.EPS,.png,.PNG,.tif,.TIF,.jpg,.JPG,.jpeg,.JPEG}}

\usepackage[utf8]{inputenc}
\usepackage[english]{babel}

\usepackage{siunitx}
\usepackage{bbm}
\newtheorem{assumption}{Assumption}

\graphicspath{{./}}

\iflatexml

\else

\paperfield{Field of the paper}
\corraddress{Nathaniel Garton, Snedecor Hall \\ 2438 Osborn Dr, Iowa State University, Ames, IA, 50010, U.S.A.}
\corremail{nmgarton@iastate.edu}
\presentadd{Snedecor Hall 2438 Osborn Dr, Iowa State University, Ames, IA, 50010, U.S.A.}
\fundinginfo{Authors Garton, Ommen, and Carriquiry were partially funded by the Center for Statistics and Applications in Forensic Evidence (CSAFE) through Cooperative Agreement \#70NANB15H176 between NIST and Iowa State University, which includes activities carried out at Carnegie Mellon University, University of California Irvine, and University of Virginia}

\papertype{Original Article}

\title{Score-based likelihood ratios to evaluate forensic pattern evidence}

\author[1]{Nathaniel  Garton}
\author[1]{Danica M. Ommen} 
\author[1]{Jarad Niemi}
\author[1]{Alicia Carriquiry}

\affil[1]{Iowa State University}

\runningauthor{Nathaniel  Garton, Danica Ommen, Jarad Niemi, Alicia Carriquiry}

\begin{document}

\maketitle
\selectlanguage{english}
\begin{abstract}

In 2016, the European Network of Forensic Science Institutes (ENFSI) published guidelines for the evaluation, interpretation and reporting of scientific evidence.  In the guidelines, ENFSI endorsed the use of the likelihood ratio (LR) as a means to represent the probative value of most types of evidence.  While computing the value of a LR is practical in several forensic disciplines, calculating an LR for pattern evidence such as fingerprints, firearm and other toolmarks is particularly challenging because standard statistical approaches are not applicable. 
Recent research suggests that machine learning algorithms can summarize a potentially large set of \emph{features} into a single score which can then be used to quantify the similarity between pattern samples. It is then possible to compute a score-based  likelihood ratio (SLR) and obtain an approximation to the value of the evidence, but research has shown that the SLR can be quite different from the LR not only in size but also in direction. We provide theoretical and empirical arguments that under reasonable assumptions, the SLR can be a practical tool for forensic evaluations.

\end{abstract}%

\section*{Introduction}
\label{intro}

Score-based likelihood ratios (SLRs) are one of the most popular methods for evaluating the strength of forensic \textit{pattern} evidence \citep{meuwly2001, gonzalezrodriguez2003,  gonzalezrodriguez2005, gonzalezrodriguez2006, neumann2009b, bolck2009, hepler2012, davis2012, bolck2015, neijmeijer2016, galbraith2017, leegwater2017, morrison2018, chen2018, hendricks2018}. Pattern evidence arises as ``the result of an impression left by a person or object" \citep{stern2017}. 
We consider the use of SLRs to determine whether two pieces of pattern evidence share a specific, known source, as opposed to a common but unknown source \citep[Chapter 3]{ommen2017}. In the \textit{specific source} framework, one piece of evidence will have an unknown source, for example a crime scene shoe print, while another piece of evidence will have a known origin, such as a suspect's shoe print. The SLR is formed by taking the ratio of distributions of a low-dimensional statistic calculated from the evidence, called a \textit{score}, under two competing hypotheses \citep{meuwly2001, davis2012, hepler2012, morrison2018}. The numerator hypothesis considers the unknown source evidence to have been generated by the suspect, while the denominator hypothesis considers the same evidence to have arisen from a source other than the suspect. The numerator hypothesis, therefore, is consistent with a prosecuting attorney's position that the suspect is guilty. Thus, we call the numerator hypothesis the \textit{prosecution} hypothesis. Alternatively, the denominator hypothesis is consistent with the defense attorney's position that the suspect is innocent and did not generate the crime scene evidence in question. Therefore, we call the denominator hypothesis the \textit{defense} hypothesis.

Scores are often a measure of dissimilarity between the piece of evidence known to come from the suspect and the crime scene evidence, however, this is not always the case \citep{morrison2018}. In the shoe print example, 
one possible score function is the Euclidean distance \citep{bolck2009, bolck2015} between the vector of pixel 
values in an image of the suspect's shoe print and an image of the crime scene shoe print. 
Using data sampled under the two competing hypotheses, the distribution of the score is modeled under both hypotheses and a ``score-based'' likelihood ratio (SLR) are \textit{estimated}. For example, one could compute Euclidean distances between images of shoe prints created 
from repeated impressions of the suspect's shoe. This would be used to model the score under the prosecution 
hypothesis. One could similarly compute Euclidean distances between images of the suspect's shoe prints 
and images of shoe prints taken from impressions of other sets of shoes. These distances could then 
be used to model the score distributions under the defense hypothesis. Examples of how score distributions are constructed 
can be found in \citet{davis2012, hepler2012}. There is a pair of implicitly 
assumed data generating models for the raw data, which are unknown but can be sampled from. 
Sampled data may come from actual casework or from other data collection efforts to build up reference databases \citep{meuwly2001}. 
For any data generating distributions under the two competing hypotheses,
 there exists (under the technical condition that the score is a measurable function) an almost surely unique SLR. 
This SLR is rarely, if ever, known in practice, and thus SLRs can only be estimated \citep{meuwly2001, gonzalezrodriguez2006}.
 However, because score distributions are low dimensional, it is reasonable to assume that techniques like density estimation would be accurate.
 Recently, researchers have successfully applied ``black box" machine 
learning classification algorithms to learn the score function by optimizing some objective function (e.g. misclassification rates on a 
training data set) \citep{srihari2002, neumann2009, hare2017, carriquiry2019}. 

Unfortunately, there is a lack of principled justification for the use of SLRs in court. A 
compelling way to establish the validity of SLRs would be to compare them to the ideal \textit{likelihood ratio} (LR) that one would get if it were possible to correctly specify distributions under the prosecution and defense hypotheses for the original, highly complex data before reduction through the score function \citep{morrison2018}. We occasionally refer to this ideal LR as the true LR.
These distributions are those implicitly assumed by the data generating procedure necessary to sample scores under the competing hypotheses. Unfortunately, several authors have raised concerning issues with SLRs. \citet{neumann2020} suggest that some types of SLRs can poorly approximate these LRs in unpredictable ways. This would be especially concerning if it were possible for an SLR and LR to strongly support opposing hypotheses. \citet{hepler2012} showed that SLRs can be constructed in multiple ways which do not always agree on the strength or the directionality of the evidence. At the same time, \citet{bolck2009, bolck2015} found that, when doing MDMA tablet comparisons, SLRs tended to be more ``stable'' and $|\log SLR|$ was commonly smaller than $|\log LR|$. 

 If SLRs (or estimates of them) are to be used in court as a method of calculating the strength forensic evidence, then it is necessary to verify that SLRs will not misrepresent the strength of evidence according to the ideal, but unknowable, LR; this is the focus in our paper. We specifically assume that the prosecution hypothesis, $H_p$, is that the evidence with unknown origin was generated by the ``distribution of the known source''. In other words, the evidence of unknown origin and the evidence of known origin can be considered to be random draws from the \textit{same} probability distribution describing repeated samples from the suspect. An example of this might be that a fingerprint at a crime scene and a fingerprint collected from a suspect are both random draws of fingerprints generated by a particular finger from the suspect. Under the defense hypothesis, $H_d$, the crime scene evidence is assumed to be generated by a different distribution than that which generated the evidence from the suspect. One intuitive, but by no means defining, example of such a hypothesis is that the crime scene evidence was generated by a random draw from a ``relevant population" \citep{bolck2009, davis2012, morrison2018}. The random draw could be according to any discrete or continuous probability distribution, not necessarily uniform. For example, the alternative population could be a single alternative source. We assume only that it is possible to sample evidence from the probability distributions that match with the prosecution and defense hypotheses for how the data was physically generated. The ratio of probabilities of the observed evidence under these true, but unknowable, distributions is the quantity that we define to be the \textit{likelihood ratio}. This definition is the same as that considered in \citet[p.~3]{royall1997}. We acknowledge, however, that this may differ from other common uses of the term ``likelihood ratio" in forensic science. For example, many researchers refer to the Bayes factor for quantifying the Bayesian value of evidence as the ``likelihood ratio" \citep{lindley1977}. This is distinct from the LR in this work because we do not advocate for or against the use of a subjective Bayesian approach to probability.


In the context of DNA evidence, one may be able to confidently and accurately estimate LRs due to established biological science \citep{stern2017}. Furthermore, measurement error, or any kind of ``within-source" variation, is likely negligible for DNA evidence. This is not the case for forensic pattern evidence. \citet{stern2017} 
discusses many of the challenges that would need to be overcome to develop LRs for pattern evidence. One of the main challenges is 
defining a probability model for a given source that could accurately describe the variability of, for example, fingerprints or shoe prints across repeated 
impressions \citep{stern2017}. Therefore, we pay particular attention to within-source variability because it plays an important role in quantifying 
evidential value for pattern and impression evidence.

In this paper, we explore the degree to which any given SLR approximates the ideal LR. We first examine a small example to illustrate when and how an SLR may misrepresent the strength of forensic evidence. We see that, even in this simple example with an intuitively reasonable score, $|\log(SLR) - \log(LR)|$ is unbounded. However, we observe that empirical probabilities of a juror making different decisions depending on whether they are provided with an SLR as opposed to an LR behave reasonably. We then generalize those ideas through the development of probabilistic bounds on the LR and argue that meaningful discrepancies between an SLR and an LR are unlikely. To our knowledge, our results provide the first non-asymptotic theoretical explanation for patterns noticed in \citet{bolck2009, bolck2015} in terms of the ``stability'' and magnitude of SLRs compared to LRs. Our results apply for any measurable score function with minimal assumptions on the true data generating models, making our results widely applicable. Further, this shows that for the types of SLRs we consider, SLRs tend to underestimate the value of evidence in a predictable way.  Other types of SLRs have empirically been shown to lack this predictability \citep{neumann2020}, and it was unknown for certain whether this lack of predictability was generally true for all SLRs. We also show results from simulation studies designed to reflect more realistic settings, which corroborate our theoretical findings and the observations of \citet{bolck2009, bolck2015}. Finally, we conclude by discussing some implications of our results.

\section*{A Simple Example} \label{s:toyexample}


The example that we study uses the same data generating model as in \citet{lindley1977} and \citet{grove1980}, but we use the notation from \citet{hepler2012}. We let $X$ and $Y$ denote the evidence of known and unknown origin, respectively. The model assumes that there are two sources of variability for evidence parameterized by a within-source variance, $\sigma_w^2$, and a between-source variance, $\sigma_b^2$. 
Within-source variance can be thought of as inherent variability of repeated samples from a particular source. Between-source variance can be thought of as the variability of the noise-free characteristic that defines the source. The model we consider is defined as follows, 

\[
\begin{array}{ll}
H_p: \hfill X \sim N(\mu_x, \sigma_w^2), & Y \sim N(\mu_x, \sigma_w^2) \\
H_d: \hfill X \sim N(\mu_x, \sigma_w^2), & Y \sim N(\mu_b, \sigma_w^2 + \sigma_b^2),
\end{array}
\]

\noindent where $X \perp Y$ under both models. The intuition for this model is that each source in the population uniquely corresponds to the mean of a Gaussian distribution, and the distribution of means within the broader population is itself distributed according to $N(\mu_b, \sigma_b^2)$. Thus, under $H_d$ the generative procedure for $Y$, considered to be sampled from a random source, can be written hierarchically as $Y|\mu_y \sim N(\mu_y, \sigma_w^2)$ and $\mu_y \sim N(\mu_b, \sigma_b^2)$.


This model is the same as in \citet{lindley1977} and \citet{grove1980} although \citet{lindley1977} puts a prior on $\mu_x$. 
Intuitively, $X$ and $Y$ are independent because the source is fixed and known under $H_p$.
We are just assuming that two pieces of evidence from within that source are sampled independently. Independence would \textit{not} be reasonable in a \textit{common source} LR under $H_p$ because the source itself is considered random rather than fixed \citep[Section 3.1]{ommen2017}. 

In this simple example, the score is $s(x,y) = (x - y)^2$. Minor variations of this score have been considered in, for example, \citet{bolck2009, bolck2015}. These distributional assumptions in combination with this score result in tractable score distributions. This pair of data generating models is almost identical to those considered in the specific source scenario in \citet[Section 2.3]{neumann2020} except we assume the variances of $X$ and $Y$ are equal under $H_p$. It is straightforward to show, using normal distribution theory, that under $H_p$, $\frac{s(X,Y)}{2 \sigma_w^2} \sim \chi_1^2$, and under $H_d$, $\frac{s(X,Y)}{2 \sigma_w^2 + \sigma_b^2} \sim \chi_1^2\left(\frac{[\mu_x - \mu_b]^2}{2 \sigma_w^2 + \sigma_b^2}\right)$, where $\frac{[\mu_x - \mu_b]^2}{2 \sigma_w^2 + \sigma_b^2}$ is the non-centrality parameter. 


In general, a set of specific source models may neither require that the distribution of the unknown source evidence comes from the same family under $H_p$ and $H_d$ nor that any parameters are shared between the two distributions. Our data generating model in this example, however, assumes a shared within-source variance, $\sigma_w^2$, for each piece of evidence. One obvious cause of variability within a source is measurement error. 

\subsection*{SLRs May Be Poor Approximations to LRs}

\begin{figure}[!htbp]
\centering
\includegraphics[height = 0.375\textheight, width = 0.56\textwidth]{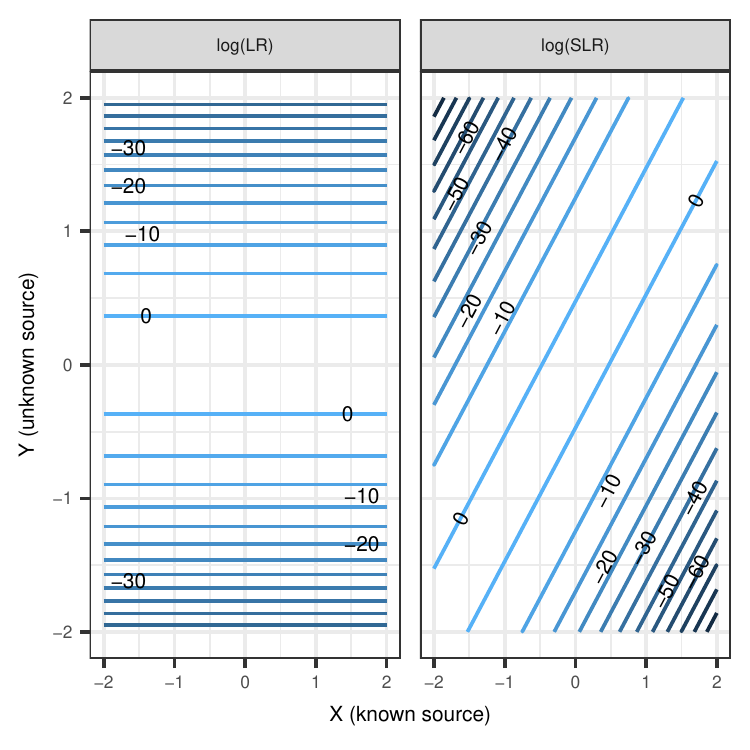}
\caption{Contour plots of $\log(LR)$ and $\log(SLR)$ at an even grid of values from -2 to 2 for $\mu_x = \mu_b = 0, \sigma_w = 0.2, \sigma_b = 1$. Contour lines are horizontal for the $\log(LR)$, but they are of the form $y = x + b$ for the $\log(SLR)$.}
\label{fig:slr_lr_contour}
\end{figure}

We first consider how the SLR compares to the LR for a grid of possible values for $X$ and $Y$. We suppose that $\mu_x = 0 =\mu_b$, so that the known source typically produces evidence commonly observed within the broader population. Further, we will fix $\sigma_b = 1$ and $\sigma_w = 0.2$. Figure \ref{fig:slr_lr_contour} compares the contour lines of $\log(LR)$ and $\log(SLR)$ on an even grid of possible $(x,y)$ values ranging from $-2$ to $2$. Comparing the LR to the SLR reveals a key difference. Because we have assumed that $X \perp Y$, the LR depends only on Y, but the SLR depends on both $X$ and $Y$. This is not surprising, but it is important to realize that this causes a potential problem. For example, if we restrict ourselves to the values of $(X,Y)$ shown in Figure \ref{fig:slr_lr_contour}, the LR achieves its minimum values at $Y = -2$ and $Y = 2$, and the SLR achieves its minimum values at $(X,Y) = (2, -2)$ and $(X,Y) = (-2, 2)$. Comparing these minimum values shows that the ratio of the LR to the SLR (or vice versa) can become very large. In this case, when $(X,Y) = (2, -2)$ the LR is roughly $3 \times 10^{19}$ times larger than the SLR.

In this example, both $X$ and $Y$ are one dimensional, the score is intuitive and simple, yet clearly large discrepancies between the LR and SLR are possible. Furthermore, while it is true that the most troubling inconsistencies between the LR and SLR occur when the observed evidence is rare with respect to the known source distribution (and thus multiplicative discrepancies of $3 \times 10^{19}$ are rare), it is not true that the known source itself is highly unusual; we have ensured that evidence generated from the known source is often very similar to that observed from the background population. Thus, it seems that such inconsistencies would be possible in most actual trials.



Perhaps the biggest problem, however, with using an SLR to approximate an LR in this example results from the differences in the contour lines between the SLR and LR. The LR contour lines are horizontal, meaning that the LR only changes with the observed values of the unknown source evidence, but the contour lines for the SLR are diagonal with slope equal to one. This is because the score distributions only depend on $(X,Y)$ through the score function. This implies that the score densities, and therefore the SLR, are constant for any given fixed score. The score is constant along lines where $y = x + b$ because when $y - x = b$ is fixed, $s(x,y) = b^2$. Because it is possible to fix the score and make $Y$ arbitrarily large or small by simply changing $X$ accordingly, we see that we can make $|\log(SLR) - \log(LR)|$ arbitrarily large. Worse, there is nothing to prevent situations where $\log(SLR)$ is positive and the $\log(LR)$ is negative, and vice versa. This implies there are situations where not only is the discrepancy between an SLR and LR large, but they are directionally inconsistent.

A final remark on the above examples is that it is impossible to determine, based solely on the score, whether there is a large discrepancy between the SLR and the LR. In this specific and seemingly reasonable case, fixing the score does not restrict the range of $Y$-values that we might obtain. This means that any given score (and consequently SLR) value can be associated with any true LR value. 

\subsection*{Probability of Large Discrepancies}
The worst problems described in the previous section involved fixing either the score 
and manipulating $(X,Y)$ (or fixing the value of $Y$ and manipulating $X$) such that their values were unlikely under either $H_p$ or $H_d$. We now show that though the probability of large discrepancies may be high, most large discrepancies will not likely affect jurors' decisions. 

\begin{figure}[!htbp]
\centering
\includegraphics[height = 0.25\textheight, width = 0.6\textwidth]{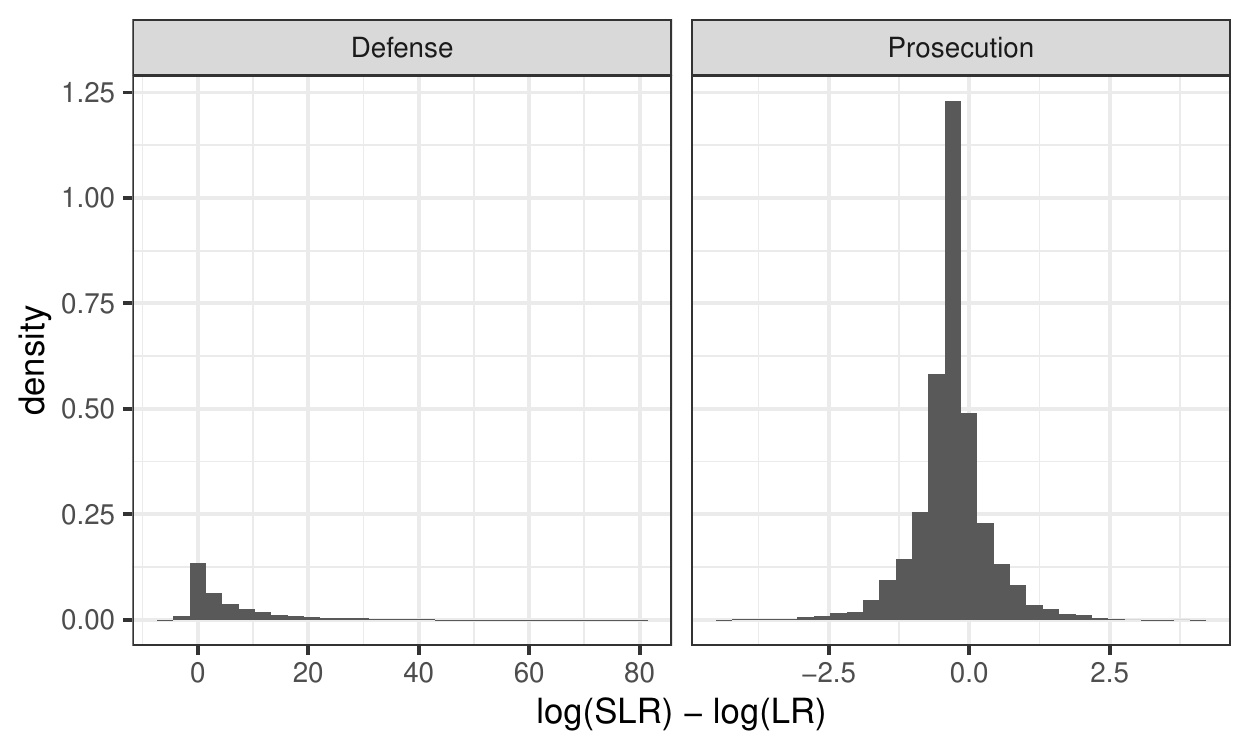}
\caption{Histograms of $\log(LR) - \log(SLR)$ generated from $5000$ samples of $(X,Y)$ values under $H_p$ (right panel) and $H_d$ (left panel).}
\label{fig:histogram_logratio}
\end{figure}

Figure \ref{fig:histogram_logratio} shows two histograms of $\log(SLR) - \log(LR)$ generated from $5000$ data sets simulated under $H_d$ and $H_p$ when $\sigma_w = 0.2, \sigma_b = 1, \mu_x = \mu_b = 0$. The left panel shows the empirical distribution of $\log(SLR) - \log(LR)$ under $H_d$, and the right panel shows the empirical distribution of the same quantity under $H_p$. We see that the distribution under $H_d$ is highly skewed right, and the smallest values that $\log(SLR) - \log(LR)$ can take are near zero. Under $H_p$, the distribution is fairly symmetric and unimodal, and most values are between $-3$ and $3$. This difference implies that the directionality and the severity of the discrepancy between the SLR and the LR may be highly dependent on whether or not $H_p$ or $H_d$ is actually true. It also shows that the probability of large discrepancies can be very high. For example, approximately $20\%$ of data sets generated under $H_d$ result in values of $\log(SLR) - \log(LR)$ larger than $10$.


\subsection*{Impact of Discrepancies on Jurors' Decisions}

Such discrepancies arguably only matter insofar as they have the potential to impact a juror's decision. With this in mind, we consider a set of bins for values of the LR and assume that a juror's decision is only impacted by the bin in which the LR falls, not its exact value.  The notion that LRs should be presented in such a way has already been suggested in, for example, \citet{nordgaard2012, enfsi}. However, we use this binning only to illustrate that the effect of common discrepancies between the SLR and the true LR on a juror's ultimate decision may often be negligible, not to advocate for any particular method for communicating the value of the SLR. We use the ranges proposed in \citet{evett2000} and similar to the proposal in \citet{marquis2016}. The scale proposed in \citet{evett2000} is as follows:  
\begin{center}
\begin{tabular}{r@{\,}l@{\,}l l}
\multicolumn{3}{c}{LR Range} & Evidence to support $H_p$ \\
\hline
$    1$ & $< LR \leq$ & $   10$ & Limited \\
$   10$ & $< LR \leq$ & $  100$ & Moderate \\
$  100$ & $< LR \leq$ & $ 1000$ & Moderately strong  \\
$ 1000$ & $< LR \leq$ & $10000$ & Strong \\
$10000$ & $< LR$      &         & Very strong.
\end{tabular}
\end{center}

\noindent 
It will be convenient to define a similar collection of sets as 

\[
\mathcal{B} \equiv \{ (0,10^{-4}), (10^{-4},10^{-3}), ..., (1,10), ..., (10^4, \infty)\}.
\]

\begin{figure}[!htbp]
\centering
\includegraphics[width = 0.7\textwidth, height = 0.35\textheight]{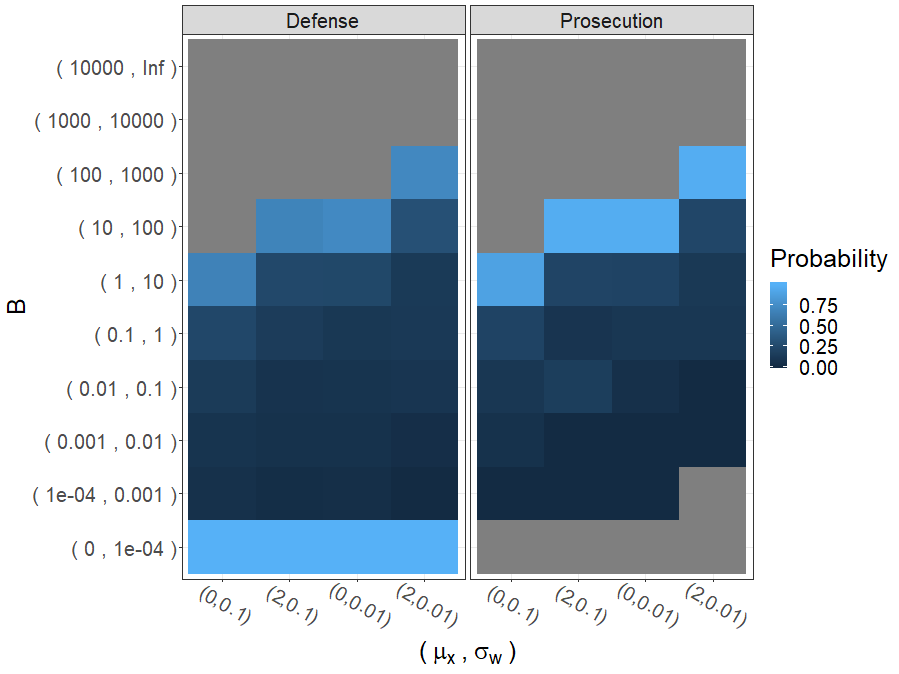}
\caption{Heatmap of empirical estimates of $P(LR \in B| SLR \in B, H_d)$ on the left and $P(LR \in B| SLR \in B, H_p)$ on the right based on $10^5$ simulated data sets. That is, for data generated under the prosecution hypothesis, this shows conditional probabilities that the LR is in the same set as the SLR. Grey areas correspond to bins in which an SLR was not observed.}
\label{fig:cond_prob_heatmap}
\end{figure}

For $B \in \mathcal{B}$, figure \ref{fig:cond_prob_heatmap} shows heatmaps of empirical conditional probabilities $P(LR \in B| SLR \in B, H_p)$ on the left and $P(LR \in B| SLR \in B, H_d)$ on the right. The probabilities were computed based on $10^{5}$ simulated observations for each parameter setting ($\mu_b = 0$ and $\sigma_b = 1$ were held fixed). The grey areas on the plots correspond to ranges of values for which no SLR was observed. We see that, under both hypotheses, only when the SLR is observed in the lowest or highest attainable bins are the probabilities of agreement between the LR and SLR close to 1.




If we examine estimates of the probabilities of agreement averaged over all sets in $\mathcal{B}$, we see that they are encouragingly large. We compute these by approximating 
\[ \sum_{B \in \mathcal{B}}{P(LR \in B| SLR \in B, H_d)P(SLR \in B|H_d)} \]
\noindent and 
\[ \sum_{B \in \mathcal{B}}{P(LR \in B| SLR \in B, H_p)P(SLR \in B|H_p)}\]
\noindent with empirical probabilities. These probabilities are shown in Table \ref{t:overall_prob_error}. As the difference between $\mu_x$ and $\mu_b$ grows or the $\sigma_b / \sigma_w$ increases, these probabilities increase as well. This means that, roughly speaking, as the difference between the known and unknown source distributions grows, the probabilities of making an error in the sense of the LR being in a different bin than the SLR decrease. 

\begin{table}[!htbp]
\centering
\begin{tabular}{rrrr}
  \hline
 $\mu_x$ & $\sigma_w$ & $H_d$ & $H_p$ \\ 
  \hline
0 & 0.1 & 0.69 & 0.87 \\ 
  2 & 0.1 & 0.93 & 0.91 \\ 
  0 & 0.01 & 0.96 & 0.93 \\ 
  2 & 0.01 & 0.99 & 0.92 \\ 
   \hline
\end{tabular}
\caption{Empirical probabilities that the LR is in the same bin as the SLR.} 
\label{t:overall_prob_error}
\end{table}

Thus, we see that while probabilities of large discrepancies may be large, the actual probability of arriving at a categorically different decision when faced with the SLR as opposed to the LR is, at worst, moderate and shrinks as there is more signal in the data to discriminate the known source from a random draw from the relevant population. This suggests that the largest and most common discrepancies occur for the most extreme values of the LR and the SLR. In the next section, we show that this pattern occurs more generally.

\section*{Probabilistic Bounds on the LR} \label{s:bounds}

By constructing probabilistic bounds on the LR conditional on the score, we provide the first theoretical justification that the patterns observed in the previous section will generalize to realistic settings. The bounds we develop are typically conservative, and only one side of each bound can be computed with only knowledge of the SLR. However, we find that these bounds provide enough insight to explain much of the behavior that we have observed up to this point. 

Denote by $p(x,y|H_i)$ the joint probability density from which the known source evidence, $X \in \mathbb{R}^{q_1}$, and the unknown source evidence, $Y \in \mathbb{R}^{q_2}$, are sampled under hypothesis $H_i$. We will use $S = s(X,Y) \in \mathbb{R}$ to denote the score random variable. We require the following assumptions for our inequalities to hold.

\begin{assumption}[Conditional Independence] \label{as:indep}
$p(x,y|H_i) = p(x|H_i)p(y|H_i)$ for $i \in \{p,d\}$.
\end{assumption}

Assumption \ref{as:indep} means that under both the prosecution and defense models, the known and unknown source evidence are generated independently  \citep{lindley1977, grove1980, aitken2004}. This assumption is reasonable for the specific source problem, but not for the common source problem. Recall that $H_p$ is equivalent to specifying that the source of $Y$ is the same as the fixed and known source of $X$. Thus, the reason that $X$ and $Y$ should be more similar under $H_p$ than $H_d$ is because they are sampled from the same marginal distribution describing repeated samples from the known source (e.g. the suspect). However, one must be careful to recognize that many papers consider common source, rather than specific source, LRs. In the common source problem, the source common to both $X$ and $Y$ is considered to be randomly chosen from a population under $H_p$, and independence between $X$ and $Y$ becomes unreasonable.

Further, our notation should not be confused with a Bayes factor, where, for example, 
\[p(x,y|H_p) = \int p(x| \theta_p, H_p)p(y| \theta_p, H_p)d\theta_p. \] Rather, 
$p(x,y|H_p)$ describes the true distribution from which data are sampled under $H_p$. Thus, \textit{if} the true distributions from which the evidence 
are sampled happen to be parametric, the true parameter values are conditioned upon. Recall that it is always necessary to assume one 
can sample from $p(x,y|H_i)$. Our results hold even when distributions are not parametric.

\begin{assumption}[Invariance of the Known Source Distribution] \label{as:xdsn}
$p(x|H_p) = p(x|H_d)$.
\end{assumption}

Assumption \ref{as:xdsn} means that regardless of whether the prosecution or defense hypothesis is true, the distribution for the known source data is the same \citep{hepler2012, bolck2015}.

\begin{assumption}[Nondegeneracy] \label{as:nondegen}
Given a fixed value $Y = y$, $S(X,y)$ is a nondegenerate random variable.
\end{assumption}

This final assumption forces the score to depend meaningfully on the known source evidence. A score function that is constant for any fixed value of $Y = y$, such as the true likelihood ratio, violates this assumption. To our knowledge, scores violating this assumption are not used. 

Under \ref{as:indep}, \ref{as:xdsn}, \ref{as:nondegen} and for $\alpha \in (1,\infty) $, we have that 

\begin{equation} \label{ineq:markov_hp}
P \left(LR \geq SLR/\alpha | s, H_p \right) \geq 1 - \frac{1}{\alpha}, 
\end{equation}

\begin{equation} \label{ineq:cauchy_hp}
P(LR \leq \alpha SLR|s, H_p) \geq \left(1 - \frac{1}{\alpha} \right)^2 \frac{SLR^{-1}}{E_{Y|s,H_d} \left[ LR^{-1} \right]},
\end{equation}

\begin{equation} \label{ineq:markov_hd}
P \left(LR \leq \alpha SLR | s, H_d \right) \geq 1 - \frac{1}{\alpha},
\end{equation}

\begin{equation} \label{ineq:cauchy_hd}
P \left( LR \geq SLR/\alpha|s, H_d \right) \geq \left(1 - \frac{1}{\alpha} \right)^2 \frac{ SLR }{E_{Y|s,H_p} \left[ LR \right]}.
\end{equation}


\noindent The derivations of these inequalities are provided in the supplementary information. Inequalities \eqref{ineq:markov_hp} and \eqref{ineq:markov_hd} are very similar to two inequalities derived in \citet[p.~7]{royall1997} for discrete probability distributions, though the derivation differs from ours. It is worth noting that in \eqref{ineq:cauchy_hp} and \eqref{ineq:cauchy_hd} the ratios multiplying $\left(1 -\frac{1}{\alpha}\right)^2$ are less than or equal to $1$. To see this, note that 

\begin{equation*}
\frac{SLR}{E_{Y|s,H_p}\left[ LR \right]} = \frac{1}{E_{Y|s,H_p}\left[ \frac{p(y|s,H_p)}{p(y|s,H_d)} \right]} \leq E_{Y|s,H_p}\left[ \frac{p(y|s,H_d)}{p(y|s,H_p)} \right] = 1,
\end{equation*}

\noindent by Jensen's inequality. A similar argument applies to $\frac{SLR^{-1}}{E_{Y|s,H_p}\left[ LR^{-1} \right]}$. This provides us with a pair of simple additional results that 

\begin{equation} \label{ineq:expectation_hp}
SLR \leq E_{Y|s,H_p}\left[ LR \right]
\end{equation}

\begin{equation} \label{ineq:expectation_hd}
SLR \geq E_{Y|s,H_d}\left[ LR^{-1} \right]^{-1}.
\end{equation}

\noindent Inequalities \eqref{ineq:expectation_hp} and \eqref{ineq:expectation_hd} can be interpreted to say that the SLR is expected to be closer to $1$ than the true LR. An additional consequence of this is that, if the LR is bounded, then the SLR must also be bounded. This is because for every score $s$, given an upper bound $M$ for the LR, $SLR \leq E_{Y|s,H_p} \left[ LR \right] \leq M $. A similar argument applies to the lower bound. Thus, if the LR is bounded, there is a limit to the degree to which the SLR can misrepresent the strength of evidence. 

Inequalities \eqref{ineq:markov_hp} and \eqref{ineq:markov_hd} explain why the SLR and the LR are likely to fall into the same bins in Figure \ref{fig:cond_prob_heatmap}. We know that, in general, there is no reason to think that the SLR is close to the LR, and so the fact that $P(LR \in B|SLR \in B, H_i)$ may be small is no surprise. However, if the SLR is sufficiently small and the defense hypothesis is true (as it was in the example), it is highly likely that both the SLR and LR will be in the same bin. For example, supposing that we observe a score such that $SLR = 10^{-5}$, then inequality \eqref{ineq:markov_hd} implies that $LR < 10^{-4}$ with at least $0.9$ probability.

A similar reasoning can be used to understand why the SLR and LR tended to fall into the same bin when the prosecution hypothesis is true. Supposing that the SLR is in the highest bin (and it often is in the example), the LR will be in at least the second highest bin at least 90\% of the time. For example, suppose $SLR \in (100, 1000)$, then $LR > 10$ with at least $0.9$ probability. And, because the LR is bounded above when all of of the data distributions are Gaussian, the LR may not be able to take values in a higher bin than the largest of the SLRs.

One practical consequence of these bounds is that if estimates of score densities are sufficiently accurate and one observes extremely large or small SLR values, it will be likely that the LR will similarly be extremely large or small, provided that large SLRs correspond to situations in which the prosecution hypothesis is true and small SLRs correspond to situations in which the defense hypothesis is true (this is a property that should be validated before any SLR method is ever used). In this case, SLRs will yield the same decisions as the ideal LR. Finally, inequalities \eqref{ineq:cauchy_hp} and \eqref{ineq:cauchy_hd} imply that if the SLR is a good approximation of the (conditional) expected value of the LR, we can establish bounds similar to those resulting from inequalities \eqref{ineq:markov_hp} and \eqref{ineq:markov_hd}. For example, assuming that $SLR^{-1} \approx E_{Y|s,H_d} \left[ LR^{-1} \right]$, we can say that $P(LR < 10 SLR | s, H_p) \gtrapprox 0.81$. 



\section*{Simulation Studies} \label{s:simulation}

\subsection*{Multivariate Normal Data}

We now consider a simulated example where the score is learned from a ``black box" machine learning classifier. We specifically consider the case where the score is a predicted class ``probability" from a trained random forest (RF). Often the ``probabilities" provided by popular implementations of random forest packages are \emph{not} directly interpretable as estimates of posterior probabilities as one might expect  \citep{pudlo2015}. We treat $H_p$ and $H_d$ as the class labels we wish to predict given the observed data, $(X,Y)$. The multivariate Gaussian example is as follows,

\[
\begin{array}{ll}
H_p: \hfill X \sim N_{5}(\mu_x, \Sigma_w), & Y \sim N_{5}(\mu_x, \Sigma_w) \\
H_d: \hfill X \sim N_{5}(\mu_x, \Sigma_w), & Y \sim N_{5}(\mu_b, \Sigma_w + \Sigma_b).
\end{array}
\]

This is essentially a multivariate version of the simple example considered in the simple example. Therefore, one should understand that each source corresponds to a particular mean for a multivariate Gaussian distribution with covariance matrix $\Sigma_w$ characterizing variation within a source. The distribution of means in the relevant population is given a Gaussian distribution with variance $\Sigma_b$. Thus, the data generating model for $Y$ under $H_d$ can be written as $Y|\mu_y \sim N_5(\mu_y, \Sigma_w)$ and $\mu_y \sim N_5(\mu_b, \Sigma_b)$.

\begin{figure}[!htbp]
\centering
\includegraphics[height = 0.35\textheight, width = 0.6\textwidth]{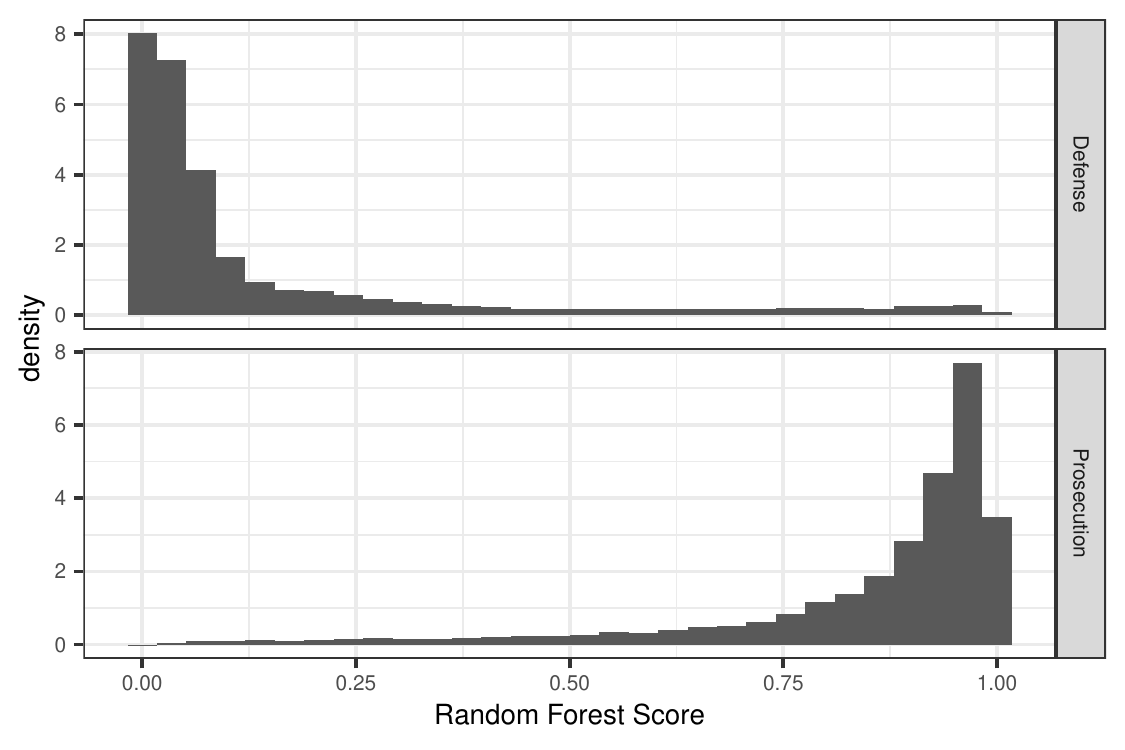}
\caption{Histograms of scores (random forest predictions) for 10000 simulated data sets under both the prosecution and defense hypothesis. Scores are generated from a random forest trained on 20000 simulated data sets, half of which correspond to the prosecution and defense hypotheses. ($\mu_x = (0.5, ..., 0.5)^\top$, $\mu_b = (0, ..., 0)^\top$, $\Sigma_w = 0.5 I_{5 \times 5}$, $\Sigma_b = I_{5 \times 5}$)}
\label{fig:rf_d5_scores}
\end{figure}

\noindent We specifically consider the case when $\mu_x = (0.5, ..., 0.5)^\top$, $\mu_b = (0, ..., 0)^\top$, $\Sigma_w = 0.5 I_{5 \times 5}$, $\Sigma_b = I_{5 \times 5}$. Figure \ref{fig:rf_d5_scores} shows histograms of 10000 scores generated under each hypothesis. The random forest was trained on 10000 data sets generated under both hypotheses which are different and independent from the data shown in these histograms. We then use kernel density estimation on the data shown in the histograms to compute score densities and SLRs. It is not always necessary to model score densities directly, if the score is an estimate of the posterior class probability. In this case one can simply multiply the estimated posterior odds by the inverse prior odds to get an estimate of the likelihood ratio. However, this is not possible with the random forest scores. Therefore, we resort to density estimation here. 

\begin{figure}[!htbp]
\centering
\includegraphics[height = 0.32\textheight, width = 0.56\textwidth]{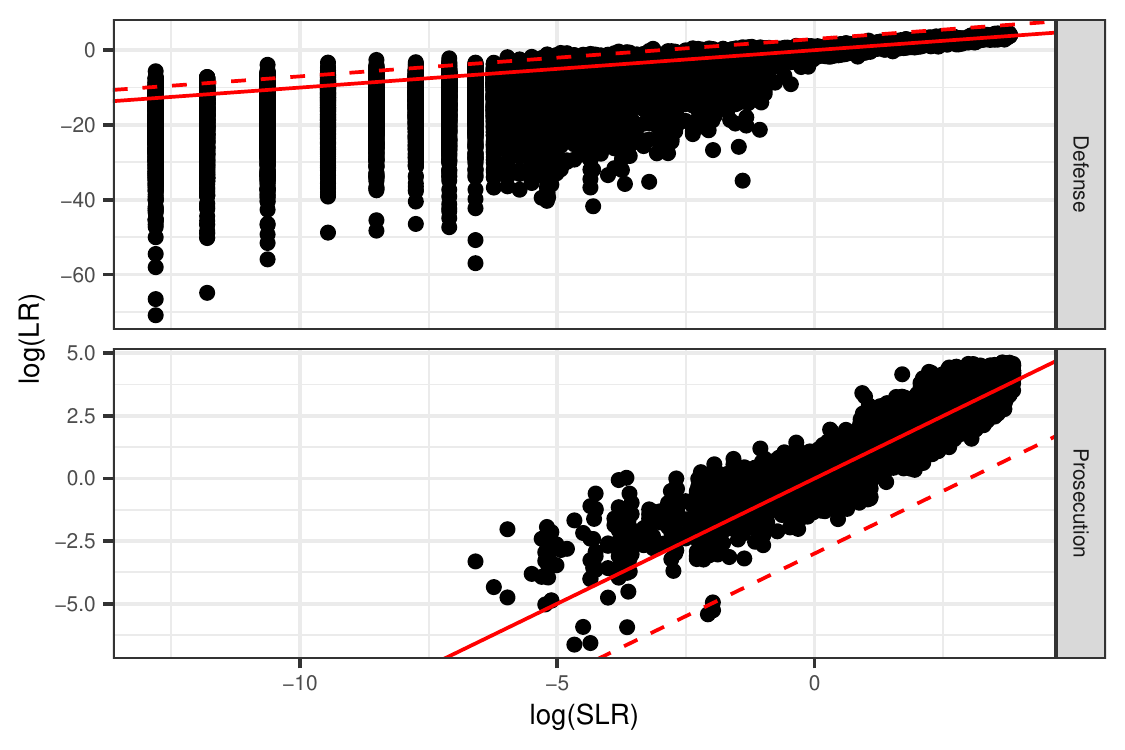}
\caption{Scatterplot of $\log(LR)$ versus $\log(SLR)$ for 10000 simulated 10 dimensional Gaussian data sets under both the prosecution and defense hypothesis. Scores are generated from a random forest trained on 20000 simulated data sets, half of which correspond to the prosecution and defense hypotheses. Score densities are estimated via kernel density estimation. The red lines correspond to what would happen if the SLR and LR were perfectly correlated and the red dashed lines correspond to $95\%$ probability bounds resulting from inequalities \eqref{ineq:markov_hp} and \eqref{ineq:markov_hd}. ($\mu_x = (0.5, ..., 0.5)^\top$, $\mu_b = (0, ..., 0)^\top$, $\Sigma_w = 0.5 I_{5 \times 5}$, $\Sigma_b = I_{5 \times 5}$)}
\label{fig:rf_n20000_d5_lr_vs_slr}
\end{figure}

Figure \ref{fig:rf_n20000_d5_lr_vs_slr} shows a scatterplot of the LR versus the SLR for 10000 simulated data sets under $H_p$ and $H_d$ (20000 total). This figure is similar to ones considered in \citet[Section 3]{neumann2020}, but \citet{neumann2020} either compare common source SLRs to specific source LRs, or they compare ``anchored" specific source SLRs to specific source LRs. An anchored SLR uses score densities that are conditioned on a value of either the known or unknown source evidence. The red line corresponds to $SLR = LR$ and the red dashed line corresponds to the conservative $95\%$ upper bound on the LR under $H_d$ and lower bound under $H_p$ resulting from inequalities \eqref{ineq:markov_hd} and \eqref{ineq:markov_hp}, respectively. Because the other set of bounds require knowledge of the conditional distribution of the LR given the score, we are unable to plot them.

We observe the same patterns in Figure \ref{fig:rf_n20000_d5_lr_vs_slr} as we did in the bivariate normal example. We see that under both hypotheses, the SLR and LR largely agree as long as the $\log(SLR)$ is not too small. Furthermore, it appears that the conditional expectation of the LR given the score is close to the SLR in this case. However, as the SLR gets smaller, which typically only happens under $H_d$, we see a wider range of possible LR values. Many LR values are much smaller than the SLR when the SLR is small itself. It is in this situation that the bounds based on inequalities \eqref{ineq:cauchy_hp} and \eqref{ineq:cauchy_hd} would be largely useless even if the required conditional expectations were known. 

We can explain some of the patterns in Figure \ref{fig:rf_n20000_d5_lr_vs_slr} using our results in the previous section. Consider when $H_d$ is true.  Inequality \eqref{ineq:markov_hd} explains why $\log(LR)$ is only rarely larger than $\log(SLR)$ and also why those discrepancies are small. Further, because the SLR itself tends to be small and $\log (LR)$ has no lower bound, the only types of large discrepancies we see are when the SLR is small but the LR is much smaller. Alternatively, when $H_p$ is true, inequality \eqref{ineq:markov_hp} explains why the $\log(LR)$ never tends to be much smaller than the $\log(SLR)$. The fact that the LR is bounded above, prohibits the large kinds of discrepancies we see under $H_p$.

We also see that the bounds resulting from inequalities \eqref{ineq:markov_hp} and \eqref{ineq:markov_hd} are typically overly conservative, with far fewer than 5\% of LRs violating the bound. Table \ref{t:markov_probabilities_mvn} provides empirical estimates of $P(LR < \alpha SLR | H_d)$ and $P(LR > SLR / \alpha |H_p)$ for six different levels of $\alpha$. The bounds based on inequalities \eqref{ineq:markov_hp} and \eqref{ineq:markov_hd} imply that all of these empirical probabilities should be greater than $1 - \frac{1}{\alpha}$, and in most cases they are much greater. 

\begin{table}[htbp!]
\centering
\begin{tabular}{r|rrrrrrr}
  \hline
$\alpha$ & 100 & 50 & 20 & 10 & 5 & 2 \\ 
\hline
  $H_d$ & 1.00 & 1.00 & 0.99 & 0.98 & 0.96 & 0.87 \\ 
  $H_p$ & 1.00 & 1.00 & 1.00 & 1.00 & 1.00 & 0.96 \\ 
   \hline
\end{tabular}
\caption{Empirical estimates of $P(LR < \alpha SLR | s, H_d)$ and $P(LR > SLR / \alpha | s, H_p)$ averaged across scores for three different levels of $\alpha$.} 
\label{t:markov_probabilities_mvn}
\end{table}

\subsection*{Beta Data Simulation}
In the previous examples, the LR, and hence the SLR, were bounded above. This was because the data distributions under $H_p$ and $H_d$ were both Gaussian and the variance under $H_d$ was larger than under $H_p$. This makes sense in our context as variability under $H_p$ is due exclusively to variation of repeated samples wtihin a source, whereas the variability of $Y$ under $H_d$ is due both to within-source variability and variability between different sources. The consequence of this was that large discrepancies between the SLR and the LR tended to occur only under $H_d$. We now provide an example where large discrepancies are possible under both hypotheses. We consider the following pair of models,  

\[
\begin{array}{ll}
H_p: \hfill X_i \stackrel{iid}{\sim} Beta(\alpha_x, \beta_x), & Y_i \stackrel{iid}{\sim} Beta(\alpha_x, \beta_x) \\
H_d: \hfill X_i \stackrel{iid}{\sim} Beta(\alpha_x, \beta_x), & Y_i \stackrel{iid}{\sim} Beta(\alpha_y, \beta_y),
\end{array}
\]

\noindent where $i = 1,...,5$. We specifically consider the case where $(\alpha_x, \beta_x) = (2,1)$ and $(\alpha_y, \beta_y) = (2,1)$.

\noindent One way to conceptualize this model is to assume, again, that for some type of forensic evidence in $(0,1)^5$ each source has a unique set of parameters $(\alpha_i, \beta_i)$. Those parameters may vary over the population according to some probability distribution. In this example, however, we assume that only two possible sources could have generated the evidence of unknown origin. Thus, only two possible sets of parameters, $(\alpha_x, \beta_x)$ or $(\alpha_x, \beta_y)$, are possible. We assume we are able to sample data from both sources.

\begin{figure}[!htbp]
\centering
\includegraphics[height = 0.35\textheight, width = 0.6\textwidth]{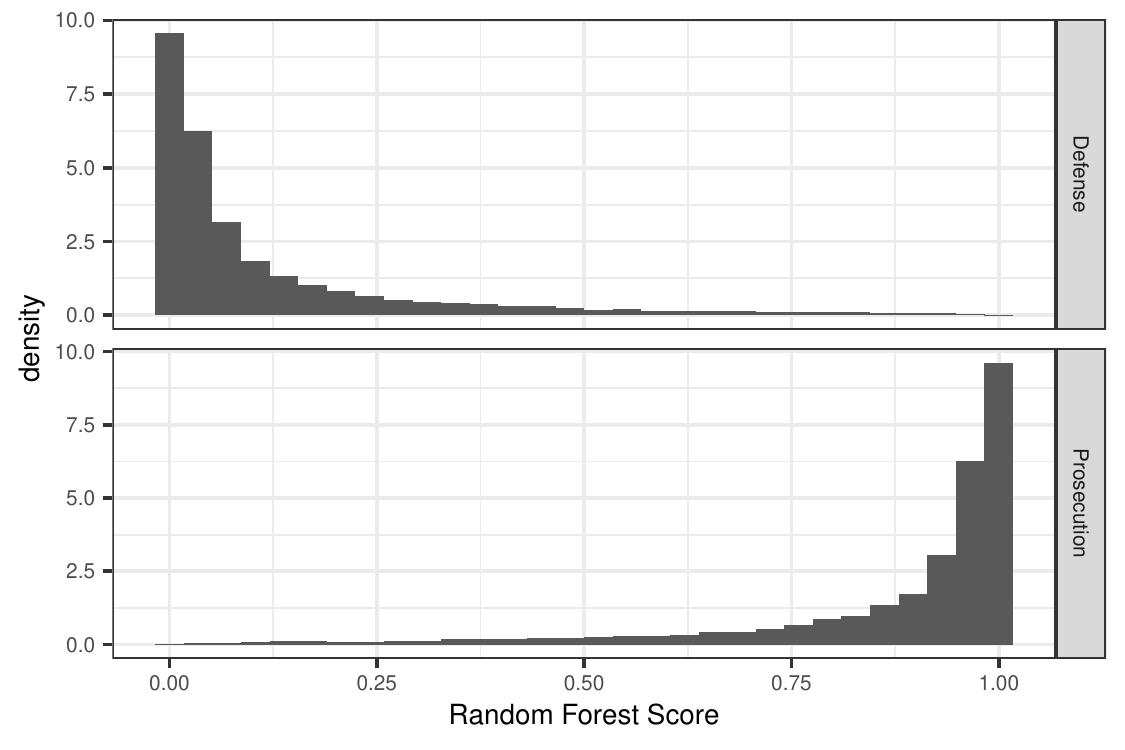}
\caption{Histograms of scores (random forest predictions) for 10000 simulated 10 dimensional Beta distributed data sets under both the prosecution and defense hypothesis. Scores are generated from a random forest trained on 20000 simulated data sets, half of which correspond to the prosecution and defense hypotheses. ($(\alpha_x, \beta_x) = (2,1)$ and $(\alpha_y, \beta_y) = (2,1)$)}
\label{fig:6}
\end{figure}

\begin{figure}[!htbp]
\centering
\includegraphics[height = 0.35\textheight, width = 0.6\textwidth]{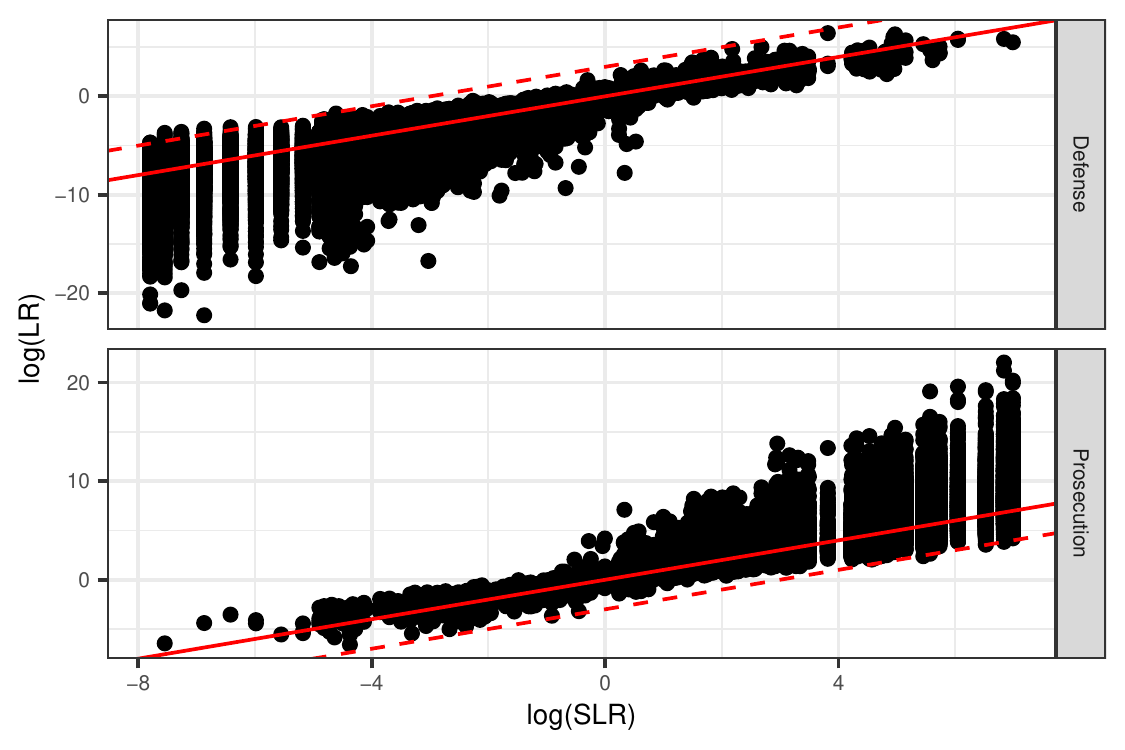}
\caption{Scatterplot of $\log(LR)$ versus $\log(SLR)$ for 10000 simulated 10 dimentional Beta distributed data sets under both the prosecution and defense hypothesis. Scores are generated from a random forest trained on 20000 simulated data sets, half of which correspond to the prosecution and defense hypotheses. Score densities are estimated via kernel density estimation. The red lines correspond to what would happen if the SLR and LR were perfectly correlated and the red dashed lines correspond to $95\%$ probability bounds resulting from inequalities \eqref{ineq:markov_hp} and \eqref{ineq:markov_hd}. ($(\alpha_x, \beta_x) = (2,1)$ and $(\alpha_y, \beta_y) = (2,1)$)}
\label{fig:7}
\end{figure}

 Figure \ref{fig:6} shows score histograms and Figure \ref{fig:7} shows scatterplots of the LR vs the SLR. Patterns in the histograms and scatterplots in this example are similar to those in the multivariate normal example. One major difference here is that the score distributions are more peaked near 1 when $H_p$ is true and near 0 when $H_d$ is true. The second major difference is that the conditional distribution of the LR given the SLR tends to be skewed both when the SLR is small and $H_d$ is true and when the SLR is large and $H_p$ is true. We still see that the bounds from inequalities \eqref{ineq:markov_hp} and \eqref{ineq:markov_hd} hold, but the bounds from inequalities \eqref{ineq:cauchy_hp} and \eqref{ineq:cauchy_hd} would not be very useful outside of $-4 < \log(SLR) < 4$. 

\section*{Discussion} \label{s:discussion}

Research, including our simple bivariate normal example, has indicated that SLRs need not always be close approximations to the true LR. The requirement, however, that the SLR be close to the LR with probability 1 is stronger than we believe is reasonable to hope for. We have shown instead that for a typical set of statistical hypotheses considered in forensic science, it is possible to establish probabilistic bounds on the LR given an observed score. These bounds are, perhaps, too loose to be used to construct interval estimates of the LR in court, but by showing that common discrepancies between SLRs and LRs are likely to be inconsequential, they support the use of SLRs as decision aids in court when LRs are unavailable.

Our theoretical results and simulation studies suggest that the largest and most common discrepancies between SLRs and LRs occur when the SLR is either very large, but the LR is much larger, or the SLR is very small, but the LR is much smaller. Among possible discrepancies, these are arguably the least troubling because SLRs are conservative -- reflecting the fact that information has been lost. This confirms that similar empirical results observed in the literature are quite general \citep{bolck2009, bolck2015}. Furthermore, these types of discrepancies will only very rarely involve large directional inconsistencies between the SLR and LR.

Our simulations involved data that was relatively low in dimension as compared to what is typically encountered in practice. It quickly becomes computationally prohibitive to accurately model the tails of the score distributions nonparametrically when the original data is high dimensional, and so we presented no such high dimensional experiments here. However, we derive in the supplementary material the following results

\begin{align}
\mathcal{D}(p(y|H_p) || p(y||H_d)) &\geq \mathcal{D}(p(s|H_p)||p(s|H_d)) \\
\mathcal{D}(p(y|H_d) || p(y||H_p)) &\geq \mathcal{D}(p(s|H_d)||p(s|H_p)),
\end{align}

\noindent where $\mathcal{D}(p(x)||q(x)) \equiv \int{\log \frac{p(x)}{q(x)}p(x) dx}$ is the Kullback-Leibler (KL) divergence between distributions $P$ and $Q$ having densities $p$ and $q$, respectively. A more general version of this result was proved in \citet{kullback1951}. The KL divergence is a measure of discrepancy between two probability distributions. Larger values of which intuitively imply that larger values of the LR under $H_p$ are common and smaller values of the LR under $H_d$ are common. 

As the data dimension increases, one would expect the KL divergence on the left hand side of the above inequalities to grow. The behavior of the right hand side, however, is not obvious. Even if the right hand side grows, it may do so slower than the left, and so the above bounds become looser and looser. This would result in the same relative behavior of the LR and the SLR shown thus far, but would likely become more extreme. 

Readers familiar with approximate Bayesian computation (ABC) \citep{tavare1997, pritchard1999, beaumont2010a, lopes2010} may recognize that, like ABC, using a score-based approach in our forensic context involves replacing intractable full data likelihoods with likelihoods based on summary statistics. Therefore, recent discussions surrounding theoretical issues encountered when using ABC for model selection \citep{robert2011, barnes2011, marin2013} are relevant to the forensic science literature. \citet{robert2011} showed that using summary statistics to do model selection in ABC can result in inconsistent (in the data sample size) Bayes factors when the statistics being used are not jointly sufficient for the model and model parameters. More importantly for our purposes, they showed that in certain situations, the discrepancy between the true Bayes factor and the approximation based on summary statistics is equivalent to a ratio of probability densities with sample size of order $n$, the number of random variables describing the observed data. The problems identified by \citet{robert2011} are fundamentally the same as those that arise when comparing an SLR to the ideal LR. However, we have shown that for the specific source problem and for a finite sample size, these discrepancies have appealing statistical properties.

Unfortunately, one of our sets of bounds involves expectations based on the conditional distribution of the data given the score, which are unavailable in practical settings. One might worry, then, about cases when the SLR is very large but the defense hypothesis is true. In this case, it is practically impossible to use our lower bound on the LR, and our experiments suggest assuming $SLR^{-1} \approx E_{Y|s,H_d} \left[ LR^{-1} \right]$ will likely be unjustifiable. Thus we can say nothing about how small the true LR might be. However, we note that 

\begin{equation}
P(LR < SLR / \alpha , SLR > \beta | H_d) \leq P(SLR > \beta | H_d).
\end{equation}

\noindent It is possible to estimate $P(SLR > \beta | H_d)$, and for large $\beta$ this probability should be very small in the first place. So while it might not be possible to provide a lower bound on the LR in this situation, we can verify that such occasions are rare to begin with. 


As a practical note, while the statistical hypotheses and assumptions we have utilized seem reasonable, score distributions may not always be generated in an appropriate way as to make the above results directly applicable. To elaborate, many score distributions are generated using samples that are necessarily dependent. One example of this might be looking at scores for all pairwise comparisons of two shoeprint images created from a suspect's shoe to generate samples from $p(s|H_p)$. Using this empirical score distribution to estimate both probability densities and probabilities of the form $P(SLR \in B|H_i)$ requires some additional assumptions and further investigation.

%

\section*{Acknowledgement}
This work was partially funded by the Center for Statistics and Applications in Forensic Evidence (CSAFE) through Cooperative Agreement \#70NANB15H176 between NIST and Iowa State University, which includes activities carried out at Carnegie Mellon University, University of California Irvine, and University of Virginia.

We are also grateful to Cedric Neumann for providing us with several additional references that we had initially missed.

\selectlanguage{english}
\bibliography{slr_references%
}

\begin{thebibliography}{39}
\providecommand{\natexlab}[1]{#1}
\providecommand{\url}[1]{\texttt{#1}}
\expandafter\ifx\csname urlstyle\endcsname\relax
  \providecommand{\doi}[1]{doi: #1}\else
  \providecommand{\doi}{doi: \begingroup \urlstyle{rm}\Url}\fi

\bibitem[Aitken and Lucy(2004)]{aitken2004}
Colin~GG Aitken and David Lucy.
\newblock Evaluation of trace evidence in the form of multivariate data.
\newblock \emph{Journal of the Royal Statistical Society: Series C (Applied
  Statistics)}, 53\penalty0 (1):\penalty0 109--122, 2004.

\bibitem[Barnes et~al.(2011)Barnes, Filippi, Stumpf, and Thorne]{barnes2011}
Chris Barnes, Sarah Filippi, Michael Stumpf, and Tom Thorne.
\newblock Considerate approaches to achieving sufficiency for {ABC} model
  selection.
\newblock \emph{Nature Precedings}, 2011.
\newblock \doi{10.1038/npre.2011.5952.1}.

\bibitem[Beaumont(2010)]{beaumont2010a}
Mark~A. Beaumont.
\newblock Approximate {B}ayesian computation in evolution and ecology.
\newblock \emph{Annual Review of Ecology, Evolution, and Systematics},
  41\penalty0 (1):\penalty0 379--406, 2010.
\newblock \doi{10.1146/annurev-ecolsys-102209-144621}.
\newblock URL \url{https://doi.org/10.1146/annurev-ecolsys-102209-144621}.

\bibitem[Bolck et~al.(2009)Bolck, Weyermann, Dujourdy, Esseiva, and van~den
  Berg]{bolck2009}
Annabel Bolck, Céline Weyermann, Laurence Dujourdy, Pierre Esseiva, and Jorrit
  van~den Berg.
\newblock Different likelihood ratio approaches to evaluate the strength of
  evidence of {MDMA} tablet comparisons.
\newblock \emph{Forensic Science International}, 191\penalty0 (1):\penalty0 42
  -- 51, 2009.
\newblock ISSN 0379-0738.
\newblock \doi{https://doi.org/10.1016/j.forsciint.2009.06.006}.
\newblock URL
  \url{http://www.sciencedirect.com/science/article/pii/S0379073809002692}.

\bibitem[Bolck et~al.(2015)Bolck, Ni, and Lopatka]{bolck2015}
Annabel Bolck, Haifang Ni, and Martin Lopatka.
\newblock Evaluating score- and feature-based likelihood ratio models for
  multivariate continuous data: applied to forensic {MDMA} comparison.
\newblock \emph{Law, Probability and Risk}, 14\penalty0 (3):\penalty0 246--266,
  2015.
\newblock \doi{10.1093/lpr/mgv009}.

\bibitem[Carriquiry et~al.(2019)Carriquiry, Hofmann, Tai, and
  Vander{P}las]{carriquiry2019}
Alicia Carriquiry, Heike Hofmann, Xiao~Hui Tai, and Susan Vander{P}las.
\newblock Machine learning in forensic applications.
\newblock \emph{Significance}, 16\penalty0 (2):\penalty0 29--35, 2019.

\bibitem[Chen et~al.(2018)Chen, Champod, Yang, Shi, Luo, Wang, Wang, and
  Lu]{chen2018}
Xiao-Hong Chen, Christophe Champod, Xu~Yang, Shao-Pei Shi, Yi-Wen Luo, Nan
  Wang, Ya-Chen Wang, and Qi-Meng Lu.
\newblock Assessment of signature handwriting evidence via score-based
  likelihood ratio based on comparative measurement of relevant dynamic
  features.
\newblock \emph{Forensic science international}, 282:\penalty0 101—110,
  January 2018.
\newblock ISSN 0379-0738.
\newblock \doi{10.1016/j.forsciint.2017.11.022}.
\newblock URL \url{https://doi.org/10.1016/j.forsciint.2017.11.022}.

\bibitem[Davis et~al.(2012)Davis, Saunders, Hepler, and Buscaglia]{davis2012}
Linda~J. Davis, Christopher~P. Saunders, Amanda Hepler, and Jo{A}nn Buscaglia.
\newblock Using subsampling to estimate the strength of handwriting evidence
  via score-based likelihood ratios.
\newblock \emph{Forensic Science International}, 2012.
\newblock \doi{10.1016/j.forsciint.2011.09.013}.

\bibitem[{European Network of Forensic Science Institutes
  (ENFSI)}(2016)]{enfsi}
{European Network of Forensic Science Institutes (ENFSI)}.
\newblock {ENFSI} guideline for evaluate reporting in forensic science, 2016.
\newblock URL
  \url{http://enfsi.eu/wp-content/uploads/2016/09/m1_guideline.pdf}.

\bibitem[Evett et~al.(2000)Evett, Jackson, Lambert, and Mc{C}rossan]{evett2000}
IW~Evett, G~Jackson, JA~Lambert, and S~Mc{C}rossan.
\newblock The impact of the principles of evidence interpretation on the
  structure and content of statements.
\newblock \emph{Science \& Justice}, 40\penalty0 (4):\penalty0 233--239,
  October 2000.
\newblock URL \url{https://doi.org/10.1016/S1355-0306(00)71993-9}.

\bibitem[Galbraith and Smyth(2017)]{galbraith2017}
Christopher Galbraith and Padhraic Smyth.
\newblock Analyzing user-event data using score-based likelihood ratios with
  marked point processes.
\newblock \emph{Digital Investigation}, 22:\penalty0 S106 -- S114, 2017.
\newblock ISSN 1742-2876.
\newblock \doi{https://doi.org/10.1016/j.diin.2017.06.009}.
\newblock URL
  \url{http://www.sciencedirect.com/science/article/pii/S1742287617301962}.

\bibitem[Gonzalez-Rodriguez et~al.(2003)Gonzalez-Rodriguez, Fierrez-Aguilar,
  and Ortega-Garcia]{gonzalezrodriguez2003}
J.~Gonzalez-Rodriguez, J.~Fierrez-Aguilar, and J.~Ortega-Garcia.
\newblock Forensic identification reporting using automatic speaker recognition
  systems.
\newblock In \emph{2003 IEEE International Conference on Acoustics, Speech, and
  Signal Processing, 2003. Proceedings. (ICASSP '03).}, volume~2, 2003.
\newblock \doi{10.1109/ICASSP.2003.1202302}.

\bibitem[Gonzalez-Rodriguez et~al.(2005)Gonzalez-Rodriguez, Fierrez-Aguilar,
  Ramos-Castro, and Ortega-Garcia]{gonzalezrodriguez2005}
Joaquin Gonzalez-Rodriguez, Julian Fierrez-Aguilar, Daniel Ramos-Castro, and
  Javier Ortega-Garcia.
\newblock {B}ayesian analysis of fingerprint, face and signature evidences with
  automatic biometric systems.
\newblock \emph{Forensic Science International}, 155\penalty0 (2):\penalty0 126
  -- 140, 2005.
\newblock ISSN 0379-0738.
\newblock \doi{https://doi.org/10.1016/j.forsciint.2004.11.007}.
\newblock URL
  \url{http://www.sciencedirect.com/science/article/pii/S0379073804007509}.

\bibitem[Gonzalez-Rodriguez et~al.(2006)Gonzalez-Rodriguez, Drygajlo,
  Ramos-Castro, Garcia-Gomar, and Ortega-Garcia]{gonzalezrodriguez2006}
Joaquin Gonzalez-Rodriguez, Andrzej Drygajlo, Daniel Ramos-Castro, Marta
  Garcia-Gomar, and Javier Ortega-Garcia.
\newblock Robust estimation, interpretation and assessment of likelihood ratios
  in forensic speaker recognition.
\newblock \emph{Computer Speech \& Language}, 20\penalty0 (2):\penalty0 331 --
  355, 2006.
\newblock ISSN 0885-2308.
\newblock \doi{https://doi.org/10.1016/j.csl.2005.08.005}.
\newblock URL
  \url{http://www.sciencedirect.com/science/article/pii/S0885230805000513}.
\newblock Odyssey 2004: The speaker and Language Recognition Workshop.

\bibitem[Grove(1980)]{grove1980}
Daniel~M. Grove.
\newblock The interpretation of forensic evidence using a likelihood ratio.
\newblock \emph{Biometrika}, 67\penalty0 (1):\penalty0 243--246, April 1980.

\bibitem[Hare et~al.(2017)Hare, Hofmann, and Carriquiry]{hare2017}
Eric Hare, Heike Hofmann, and Alicia Carriquiry.
\newblock Automatic matching of bullet land impressions.
\newblock \emph{The Annals of Applied Statistics}, 11\penalty0 (4):\penalty0
  2332--2356, December 2017.

\bibitem[Hendricks et~al.(2018)Hendricks, Neumann, and Saunders]{hendricks2018}
JH~Hendricks, C~Neumann, and CP~Saunders.
\newblock Quantifying the weight of fingerprint evidence using an {ROC}-based
  approximate {B}ayesian computation algorithm.
\newblock \emph{Journal of the Royal Statistical Society C}, 2018.

\bibitem[Hepler et~al.(2012)Hepler, Saunders, Davis, and Buscaglia]{hepler2012}
Amanda~B. Hepler, Christopher~P. Saunders, Linda~J. Davis, and Jo{A}nn
  Buscaglia.
\newblock Score-based likelihood ratios for handwriting evidence.
\newblock \emph{Forensic Science International}, 219:\penalty0 129--140, 2012.

\bibitem[Kullback and Leibler(1951)]{kullback1951}
S.~Kullback and R.A. Leibler.
\newblock On information and sufficiency.
\newblock \emph{Annals of Mathematical Statistics}, 22\penalty0 (1):\penalty0
  79--86, 1951.
\newblock \doi{10.1214/aoms/1177729694}.
\newblock URL \url{https://doi.org/10.1214/aoms/1177729694}.

\bibitem[Leegwater et~al.(2017)Leegwater, Meuwly, Sjerps, Vergeer, and
  Alberink]{leegwater2017}
Anna~Jeannette Leegwater, Didier Meuwly, Majan Sjerps, Peter Vergeer, and Iwo
  Alberink.
\newblock Performance study of a score-based likelihood ratio system for
  forensic fingermark comparison.
\newblock \emph{Journal of Forensic Sciences}, 62\penalty0 (3), 2017.
\newblock \doi{10.1111/1556-4029.13339}.

\bibitem[Lindley(1977)]{lindley1977}
Dennis~V. Lindley.
\newblock A problem in forensic science.
\newblock \emph{{B}iometrika}, 64\penalty0 (2):\penalty0 207--213, August 1977.

\bibitem[Lopes and Beaumont(2010)]{lopes2010}
J.S. Lopes and M.A. Beaumont.
\newblock {ABC}: A useful {B}ayesian tool for the analysis of population data.
\newblock \emph{Infection, Genetics and Evolution}, 10\penalty0 (6):\penalty0
  825--832, 2010.
\newblock \doi{10.1016/j.meegid.2009.10.010}.

\bibitem[Marin et~al.(2013)Marin, Pillai, Robert, and Rousseau]{marin2013}
Jean-Michel Marin, Natesh~S. Pillai, Christian~P. Robert, and Judith Rousseau.
\newblock Relevant statistics for {B}ayesian model choice.
\newblock \emph{Journal of the Royal Statistical Society: Series B (Statistical
  Methodology)}, 76\penalty0 (5):\penalty0 833--859, 2013.
\newblock \doi{10.1111/rssb.12056}.
\newblock URL
  \url{https://rss.onlinelibrary.wiley.com/doi/abs/10.1111/rssb.12056}.

\bibitem[Marquis et~al.(2016)Marquis, Biedermann, Cadola, Champod, Gueissaz,
  Massonet, Mazzella, Taroni, and Hicks]{marquis2016}
Raymond Marquis, Alex Biedermann, Liv Cadola, Christophe Champod, Line
  Gueissaz, Genevi{\`e}ve Massonet, Williams~David Mazzella, Franco Taroni, and
  Tacha Hicks.
\newblock Discussion on how to implement a verbal scale in a forensic
  laboratory: Benefits, pitfalls and suggestions to avoid misunderstandings.
\newblock \emph{Science and Justice}, 56\penalty0 (5):\penalty0 364--370,
  September 2016.
\newblock URL \url{https://doi.org/10.1016/j.scijus.2016.05.009}.

\bibitem[Meuwly and Drygajlo(2001)]{meuwly2001}
Didier Meuwly and Andrzej Drygajlo.
\newblock Forensic speaker recognition based on a bayesian framework and
  gaussian mixture modelling (gmm).
\newblock In \emph{2001: A Speaker Odyssey-The Speaker Recognition Workshop},
  2001.

\bibitem[Morrison and Enzinger(2018)]{morrison2018}
Geoffrey~Stewart Morrison and Ewald Enzinger.
\newblock Score based procedures for the calculation of forensic likelihood
  ratios - scores should take account of both similarity and typicality.
\newblock \emph{Science and Justice}, 58:\penalty0 47--58, 2018.

\bibitem[Neijmeijer(2016)]{neijmeijer2016}
Ren\`{e} Neijmeijer.
\newblock Assessing performance of score-based likelihood ratio methods for
  forensic data.
\newblock Master's thesis, Leiden University, 2016.
\newblock URL
  \url{https://openaccess.leidenuniv.nl/bitstream/handle/1887/44582/Neijmeijer%2C%20Ren%C3%A9-s1436643-MA%20Thesis%20MS-2016.pdf?sequence=1}.

\bibitem[Neumann and Ausdemore(2020)]{neumann2020}
C.~Neumann and A.~Ausdemore.
\newblock Defence against the modern arts: the curse of statistics
  {``Score-based likelihood ratios"}.
\newblock \emph{Law, Probability and Risk}, 2020.

\bibitem[Neumann and Margot(2009{\natexlab{a}})]{neumann2009}
Cedric Neumann and Pierre Margot.
\newblock New perspectives in the use of ink evidence in forensic science: Part
  {II}. development and testing of mathematical algorithms for the automatic
  comparison of ink samples analysed by {HPTLC}.
\newblock \emph{Forensic Science International}, 185\penalty0 (1-3):\penalty0
  38--50, 2009{\natexlab{a}}.

\bibitem[Neumann and Margot(2009{\natexlab{b}})]{neumann2009b}
Cedric Neumann and Pierre Margot.
\newblock New perspectives in the use of ink evidence in forensic science part
  {III}: Operational applications and evaluation.
\newblock \emph{Forensic Science International}, 2009{\natexlab{b}}.

\bibitem[Nordgaard and Rasmusson(2012)]{nordgaard2012}
Anders Nordgaard and Birgitta Rasmusson.
\newblock The likelihood ratio as value of evidence - more than a question of
  numbers.
\newblock \emph{Law, Probability and Risk}, 11\penalty0 (4):\penalty0 303--315,
  July 2012.
\newblock \doi{10.1093/lpr/mgs019}.

\bibitem[Ommen(2017)]{ommen2017}
Danica Ommen.
\newblock \emph{Approximate statistical solutions to the forensic
  identification of source problem}.
\newblock PhD thesis, South Dakota State University, 2017.
\newblock URL
  \url{https://openprairie.sdstate.edu/cgi/viewcontent.cgi?referer=https://scholar.google.com/&httpsredir=1&article=2780&context=etd}.

\bibitem[Pritchard et~al.(1999)Pritchard, Seielstad, Perez-Lezaun, and
  Feldman]{pritchard1999}
Jonathon~K. Pritchard, Mark~T. Seielstad, Anna Perez-Lezaun, and Marcus~W.
  Feldman.
\newblock Population growth of human {Y} chromosomes: a study of {Y} chromosome
  microsatellites.
\newblock \emph{Molecular biology and evolution}, 16\penalty0 (12):\penalty0
  1791--1798, 1999.
\newblock \doi{10.1093/oxfordjournals.molbev.a026091}.

\bibitem[Pudlo et~al.(2015)Pudlo, Marin, Estoup, Cornuet, Gautier, and
  Robert]{pudlo2015}
Pierre Pudlo, Jean-Michel Marin, Arnaud Estoup, Jean-Marie Cornuet, Mathieu
  Gautier, and Christian~P Robert.
\newblock Reliable {ABC} model choice via random forests.
\newblock \emph{Bioinformatics}, 32\penalty0 (6):\penalty0 859--866, 2015.

\bibitem[Robert et~al.(2011)Robert, Cornuet, Marin, and Pillai]{robert2011}
Christian~P. Robert, Jean-Marie Cornuet, Jean-Michel Marin, and Natesh~S.
  Pillai.
\newblock Lack of confidence in approximate {B}ayesian computation of model
  choice.
\newblock \emph{Proceedings of the National Academy of Sciences}, 108\penalty0
  (37):\penalty0 15112--15117, September 2011.
\newblock URL \url{https://doi.org/10.1073/pnas.1102900108}.

\bibitem[Royall(1997)]{royall1997}
Richard~M. Royall.
\newblock \emph{Statistical evidence: a likelihood paradigm}.
\newblock Chapman \& Hall, 1997.

\bibitem[Srihari et~al.(2002)Srihari, Cha, Arora, and Lee]{srihari2002}
Sargur~N. Srihari, Sung-Hyuk Cha, Hina Arora, and Sangjik Lee.
\newblock Individuality of handwriting.
\newblock \emph{Journal of Forensic Science}, 47\penalty0 (4), 2002.

\bibitem[Stern(2017)]{stern2017}
Hal~S. Stern.
\newblock Statistical issues in forensic science.
\newblock \emph{Annual Review of Statistics and Its Applications}, 4:\penalty0
  2, 2017.

\bibitem[Tavar\'{e} et~al.(1997)Tavar\'{e}, Balding, Griffiths, and
  Donnelly]{tavare1997}
Simon Tavar\'{e}, David~J. Balding, R.C. Griffiths, and Peter Donnelly.
\newblock Inferring coalescence times from {DNA} sequence data.
\newblock \emph{Genetics}, 145\penalty0 (2):\penalty0 505--518, 1997.

\end{thebibliography}







\section*{Supplementary Information}
\subsection*{Derivation of Inequalities}
We now derive inequalities 1-4. First, note that assumptions 1 and 2 imply that $\frac{p(x,y|H_p)}{p(x,y|H_d)} = \frac{p(y|H_p)}{p(y|H_d)}$.

Consider the joint density functions of $(Y,s(X,Y)) = (Y,S)$ under $H_d$ and $H_p$ denoted by $p(y,s|H_d)$ and $p(y,s|H_p)$, respectively. Note that the likelihood ratio of the augmented data vector $(Y,S)$ is equal to that of the unaugmented data $Y$. This is because 
\begin{align*}
\frac{p(y,s|H_p)}{p(y,s|H_d)} &= \frac{p(s|y, H_p)p(y|H_p)}{p(s|y,H_d)p(y|H_d)} \\
&= \frac{p(s|y)p(y|H_p)}{p(s|y)p(y|H_d)} \\
&= \frac{p(y|H_p)}{p(y|H_d)}.
\end{align*}

\noindent Because $s(X,Y = y)$ is random only due to $X$, and because $p(x|H_p) = p(x|H_d)$, we have that $p(s|y, H_p) = p(s|y,H_d)$.

With this in mind, note that

\begin{align*}
\frac{p(y,x|H_p)}{p(y,x|H_d)} &= \frac{p(y|H_p)}{p(y|H_d)} \\
&= \frac{p(y,s|H_p)}{p(y,s|H_d)} \\
&= \frac{p(y|s, H_p) p(s|H_p)}{p(y|s,H_d) p(s|H_d)}.
\end{align*}


We use a simple application of Markov's inequality to show that,

\begin{align*}
P \left(LR < 1/c | s, H_p \right) &= P \left(\frac{1}{LR} > c | s, H_p \right) \\ 
&= P \left(\frac{p(y|H_d)}{p(y|H_p)} > c | s, H_p \right) \\ 
&= P\left(\frac{p(y|s, H_d)p(s|H_d)}{p(y|s, H_p)p(s|H_p)} > c | s, H_p \right) \\
&= P \left(\frac{p(y|s, H_d)}{p(y|s, H_p)} > c \frac{p(s|H_p)}{p(s|H_d)} | s, H_p \right) \\
&\leq \frac{1}{c \frac{p(s|H_p)}{p(s|H_d)}} E_{Y|s,H_p}\left[ \frac{p(y|s, H_d)}{p(y|s, H_p)} \right] \\
&= \frac{1}{c * SLR}.
\end{align*}

\noindent Equivalently, we can write

\begin{equation*}
P \left(LR > 1/c | s, H_p \right) \geq 1 - \frac{1}{c * SLR}.
\end{equation*}

\noindent Thus we find a lower bound on the probability that, under $H_p$, the LR exceeds $1/c$. Let us consider values of the form $c = \frac{\alpha}{SLR}$ where $\alpha \in (1,\infty)$. This results in inequality 1.


We can use a symmetric argument to create a probabilistic upper bound on the LR given the score under the defense hypothesis. That is,

\begin{align*}
P \left(LR > c | s, H_d \right) &= P \left(\frac{p(y|H_p)}{p(y|H_d)} > c | s, H_d \right) \\
&= P\left(\frac{p(y|s, H_p)p(s|H_p)}{p(y|s, H_d)p(s|H_d)} > c | s, H_d \right) \\
&= P \left(\frac{p(y|s, H_p)}{p(y|s, H_d)} > c \frac{p(s|H_d)}{p(s|H_p)} | s, H_d \right) \\
&\leq \frac{1}{c \frac{p(s|H_d)}{p(s|H_p)}} E_{Y|s,H_d}\left[ \frac{p(y|s, H_p)}{p(y|s, H_d)} \right] \\
&= \frac{SLR}{c}.
\end{align*}

\noindent Now, considering $c$ of the form $c = \alpha SLR$ where still $\alpha \in (1,\infty)$, we get inequality 3.


While the bounds derived from Markov's inequality help explain the relationship between an SLR and the LR, they are unhelpful in terms of giving an upper bound on the LR in the case that the prosecution's hypothesis is true or a lower bound on the LR in the case that the defense hypothesis is true. We now derive such bounds from Cauchy-Scwhartz's inequality. Unfortunately, these bounds involve incomputable quantities based on the true data densities. They also require that $E_{Y|s,H_p} \left[ LR \right] < \infty$ and $E_{Y|s,H_d} \left[ LR^{-1} \right] < \infty$ in order to be non-trivial. Define the indicator function 

\[
\mathbbm{1}\left[ x \in A \right] = \begin{cases} 
      0 & x \notin A \\
      1 & x \in A.
   \end{cases}
\]


For now, note that Cauchy-Schwartz's inequality implies 

\begin{align*}
E_{Y|s,H_d} \left[ LR \mathbbm{1}\left[ LR > c \right] \right] &= E_{Y|s,H_d} \left[ \frac{p(y|H_p)}{p(y|H_d)} \mathbbm{1}\left[ \frac{p(y|H_p)}{p(y|H_d)} > c \right] \right] \\
&\leq \begin{multlined}[t]
E_{Y|s,H_d} \left[ \left(\frac{p(y|H_p)}{p(y|H_d)}\right)^2 \right]^{1/2} \\
\times E_{Y|s,H_d} \left[ \mathbbm{1}\left[ \frac{p(y|H_p)}{p(y|H_d)} > c \right]^2 \right]^{1/2}
\end{multlined} \\
&= \begin{multlined}[t]
E_{Y|s,H_d} \left[ \left(\frac{p(y|s,H_p)p(s|H_p)}{p(y|s,H_d)p(s|H_d)}\right)^2 \right]^{1/2} \\ \times P \left( \frac{p(y|H_p)}{p(y|H_d)} > c |s,H_d \right)^{1/2}. 
\end{multlined}
\end{align*}

\noindent We will now write $\frac{p(y|H_p)}{p(y|H_d)}$ as $LR$ for notational compactness. Rearranging terms and squaring both sides we see that

\begin{align*}
P \left( LR > c|s, H_d \right) &\geq
\begin{multlined}[t] \frac{ E_{Y|s,H_d} \left[ \frac{p(y|s,H_p)}{p(y|s,H_d)} \mathbbm{1}\left[ LR > c \right] \right]^2}{ E_{Y|s,H_d} \left[ \left(\frac{p(y|s,H_p)}{p(y|s,H_d)}\right)^2 \right]} \\
 \times \frac{\left(\frac{p(s|H_p)}{p(s|H_d)}\right)^2}{\left(\frac{p(s|H_p)}{p(s|H_d)}\right)^2}
\end{multlined} \\
&=  \frac{ \left(\frac{p(s|H_p)}{p(s|H_d)}\right)E_{Y|s,H_p} \left[ \mathbbm{1}\left[ LR > c \right] \right]^2}{\left(\frac{p(s|H_p)}{p(s|H_d)}\right)  E_{Y|s,H_p} \left[ \frac{p(y|s,H_p)}{p(y|s,H_d)} \right]} \\
&= \frac{\left(\frac{p(s|H_p)}{p(s|H_d)}\right) P \left( LR > c | s, H_p \right)^2}{E_{Y|s,H_p} \left[ LR \right]} \\
&= SLR\frac{P \left(LR > c | s, H_p \right)^2}{E_{Y|s,H_p} \left[ LR \right]}.
\end{align*}

\noindent By plugging in $c = SLR/\alpha$ and using the bound from Markov's inequality, we arrive at the final bound

\begin{align*}
P \left( LR > SLR/\alpha|s, H_d \right) &\geq SLR\frac{P \left( LR > SLR/\alpha | s, H_p \right)^2}{E_{Y|s,H_p} \left[ LR \right]} \\
&\geq \left(1 - \frac{1}{\alpha} \right)^2 \frac{ SLR }{E_{Y|s,H_p} \left[ LR \right]}.
\end{align*}

In a similar way, we can use Cauchy-Schwartz Inequality on \\ 
\noindent$E_{Y|s,H_p}\left[ LR^{-1} \mathbbm{1}\left[ LR^{-1} > c \right] \right]$ to arrive at 

\begin{align*}
P(LR^{-1} > c|s, H_p) &\geq \frac{E_{Y|s,H_p}\left[ LR^{-1} \mathbbm{1}\left[ LR^{-1} > c \right] \right]^2}{E_{Y|s,H_p} \left[ LR^{-2} \right]} \\
&= \frac{E_{Y|s,H_p}\left[ \frac{p(y|s,H_d)}{p(y|s,H_p)} \mathbbm{1}\left[ LR^{-1} > c \right] \right]^2}{E_{Y|s,H_p} \left[ \left( \frac{p(y|s,H_d)}{p(y|s,H_p)} \right)^2 \right]} \\
&= SLR^{-1} \frac{P(LR < 1/c | s, H_d)^2}{E_{Y|s,H_d} \left[ LR^{-1} \right]}.
\end{align*}

\noindent Again, using the bound from Markov's inequality in the last section, we arrive at 

\begin{align*}
P(LR < \alpha SLR|s, H_p) &\geq SLR^{-1} \frac{P(LR < \alpha SLR | s, H_d)^2}{E_{Y|s,H_d} \left[ LR^{-1} \right]} \\
&= \left(1 - \frac{1}{\alpha} \right)^2 \frac{SLR^{-1}}{E_{Y|s,H_d} \left[ LR^{-1} \right]}.
\end{align*}

\subsection*{Discrepancy as a Function of Data Dimension}


The Kullback-Leibler (KL) divergence of a distribution $P$ to a distribution $Q$ is defined as 
\begin{equation*}
\mathcal{D}(p(x)||q(x)) \equiv \int{\log \frac{p(x)}{q(x)}p(x) dx},
\end{equation*}

\noindent where $p(x)$ and $q(x)$ are the probability densities. 

For the following derivations, we assume that $X$ and $Y$ have densities with respect to the Lebesgue measure and that the score function is continuous. However, the following results should generalize to discrete data as well. With assumptions 1 and 2, it is simple to show that $\mathcal{D}(p(x,y|H_p) || p(x,y|H_d)) = \mathcal{D}(p(y|H_p) || p(y|H_d)) = \mathcal{D}(p(y,s|H_p) || p(y,s|H_d))$. Furthermore, we can decompose $\mathcal{D}(p(y|H_p) || p(y||H_d)$ as 

\begin{align*} 
\mathcal{D}(p(y|H_p) &|| p(y||H_d) = \int{\log \frac{p(y|H_p)}{p(y|H_d)} p(y|H_p) dy} \\
&= \int {\int{\log \frac{p(y,s|H_p)}{p(y,s|H_d)} p(y,s|H_p) dy} ds} \\
&= \int { \left( \int{\log \frac{p(y|s,H_p)p(s|H_p)}{p(y|s,H_d)p(s|H_d)} p(y,s|H_p) dy} \right) ds} \\
&= \begin{multlined}[t] \int{ \left( \int{\log \frac{p(y|s,H_p)}{p(y|s,H_d)} p(y|s,H_p) dy} \right) p(s|H_p) ds} \\
 + \int{ \left( \int{\log \frac{p(s|H_p)}{p(s|H_d)} p(y|s,H_p) dy} \right) p(s|H_p) ds} \end{multlined} \\
&= \begin{multlined}[t] E_{S|H_p}\left[ \mathcal{D}(p(y|s,H_p)||p(y|s,H_d)) \right] \\
+ \mathcal{D}(p(s|H_p)||p(s|H_d)) \label{eq:kl_decomp}. \end{multlined}
\end{align*}

\noindent Note that because $E_{S|H_p}\left[ \mathcal{D}(p(y|s,H_p)||p(y|s,H_d)) \right]$ is nonnegative, $\mathcal{D}(p(s|H_p)||p(s|H_d))$ is a lower bound for the KL divergence of the raw data. 


\end{document}